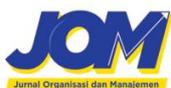

**Jurnal Organisasi dan Manajemen**
Journal Homepage : http://jurnal.ut.ac.id/index.php/JOM

# Strategic ESG-Driven Human Resource Practices: Transforming Employee Management for Sustainable Organizational Growth

Darul Wiyono[1*], Deshinta Arrova Dewi[2], Ema Ambiapuri[3], Nur Aini Parwitasari[3], Deni Supardi Hambali[3]

1. Department of Administrative Management, Akademi Sekretari dan Manajemen Ariyanti, Indonesia
2. Faculty of Data Science and Information Technology, INTI International University, Malaysia
3. Department of Secretary, Akademi Sekretari dan Manajemen Ariyanti, Indonesia
*Corresponding author e-mail: darulwiyono96@ariyanti.ac.id

**Abstract**

**Purpose** - This research explores the impact of Environmental, Social, and Governance (ESG) practices on employee performance and well-being in private higher education institutions in Bandung, West Java. It seeks to provide insights into how effective ESG integration can enhance organizational performance and employee satisfaction.
**Methodology** - A quantitative approach employing Partial Least Squares Structural Equation Modeling (PLS-SEM) analyzed the relationships among constructs: environmental practices, social practices, governance practices, and the dependent variable, employee performance and well-being. Data were collected from 270 respondents through stratified random sampling across various administrative roles in 138 private higher education institutions.
**Findings**- The results showed that environmental practices positively impact employee performance and well-being, with social practices also contributing. Governance practices mediate and amplify these effects. These findings emphasize the importance of integrating sustainable practices into organizational strategies to improve employee outcomes and overall institutional performance.
**Originality**—This study explores the impact of ESG practices on employee performance in private higher education institutions, focusing on sustainability to enhance engagement and productivity. Unlike previous research, which focused primarily on corporate sectors or public universities, it emphasizes the unique challenges of private institutions, particularly in Bandung, offering new insights into governance practices as mediators.

## 1. Introduction

Human resource practices are pivotal in enhancing organizational performance by fostering employee satisfaction, engagement, and productivity. These practices are essential for creating a conducive work environment that aligns with organizational goals. In recent years, integrating Environmental, Social, and Governance (ESG) principles has gained significant attention as a transformative approach in human resource management (Zhang et al., 2024). ESG focuses on environmental stewardship, social responsibility, and ethical governance, making it a cornerstone for modern administrative strategies. The adoption of ESG principles not only meets regulatory compliance but also addresses the increasing concerns of stakeholders who demand sustainable and responsible management practices (Agapova & Garanina, 2024)

Several determinants influence human resource practices, including organizational culture, leadership, technological advancements, and stakeholder expectations. Existing literature highlights the critical role of these factors in shaping effective administrative policies (Leal Filho et al., 2023; Al-Alawneh et al., 2024). The evolving landscape of higher education demands innovative approaches to ensure institutions remain competitive while fulfilling their social and environmental responsibilities. For example, institutions that embed sustainability into their mission and vision report enhanced employee engagement and institutional performance (Sanches et al., 2023). Despite these advancements, there remains a significant gap in understanding how ESG principles specifically impact human resource practices within the context of higher education institutions (Mokski et al., 2023; García-Rodríguez & Gutiérrez-Taño, 2024)

While existing studies have explored the impact of ESG principles on corporate performance, the extent to which these principles influence human resource practices in resource-constrained environments, such as private universities, remains underexplored. This gap is particularly significant given private universities' unique challenges in balancing financial constraints, regulatory requirements, and stakeholder expectations (Habib & Mourad, 2023). Furthermore, the debate continues whether the primary motivation for ESG adoption lies in achieving compliance or fostering long-term sustainability and employee well-being (Buser et al., 2024).

Integrating ESG principles into human resource practices may act as a moderating factor that amplifies their effectiveness. For instance, ESG-driven policies can mitigate workplace stress, enhance employee retention, and promote a healthy organizational climate (Ronalter et al., 2023). Moderation is critical to understanding the extent to which ESG principles influence the relationship between administrative practices and employee productivity. By investigating these interactions, this study provides a nuanced understanding of the mechanisms through which ESG impacts institutional performance.

This research contributes to the existing body of literature by addressing the gap in understanding ESG's role in higher education administration. Academically, it enriches theoretical frameworks related to sustainable human resource management and organizational performance. Socially, it highlights the importance of ESG in fostering employee well-being and a productive work environment. From a policy perspective, the findings offer valuable recommendations for designing regulatory frameworks that support sustainable administrative practices in higher education institutions (Adeusi et al., 2024; Al-Alawneh et al., 2024). Moreover, the research underscores the strategic importance of integrating ESG principles within organizational management to align with stakeholder expectations and improve institutional effectiveness (Said et al., 2023). Corporate Social Responsibility (CSR) disclosure has also gained attention as an essential factor in evaluating an organization's commitment to sustainable practices (Wulolo & Rahmawati, 2017). In line with this, strategies and innovations for enhancing sustainable performance play a crucial role in advancing long-term organizational success (Cuandra & Candy, 2024).

This study's framework is designed to explain the relationship between environmental, social, and governance practices and employee performance and well-being in private universities in Bandung. This framework includes influential variables, namely Environmental Practices ($X_1$), Social Practices ($X_2$), and Governance Practices ($Z$), and their impact on employee performance and well-being (Y). By mapping the interactions between these variables, this research aims to provide a more comprehensive insight into how implementing ESG practices can enhance institutional performance and staff well-being.

Governance practices (Z) are positioned as mediators in this framework based on a strong theoretical foundation. Governance bridges environmental and social initiatives to tangible outcomes, ensuring policies are effectively implemented and aligned with institutional goals. According to stakeholder theory, robust governance mechanisms are essential for balancing the diverse interests of stakeholders, including employees, students, and external partners. Moreover, governance ensures accountability and transparency, which are critical for fostering organizational trust and commitment. Research by Jain et al. (2024) highlights that governance practices mediate the relationship between sustainability efforts and organizational performance by creating a structured and cohesive approach to policy implementation. Similarly, Andrey (2023) emphasizes that governance provides the necessary oversight and strategic direction to translate environmental and social practices into meaningful employee performance and well-being improvements.

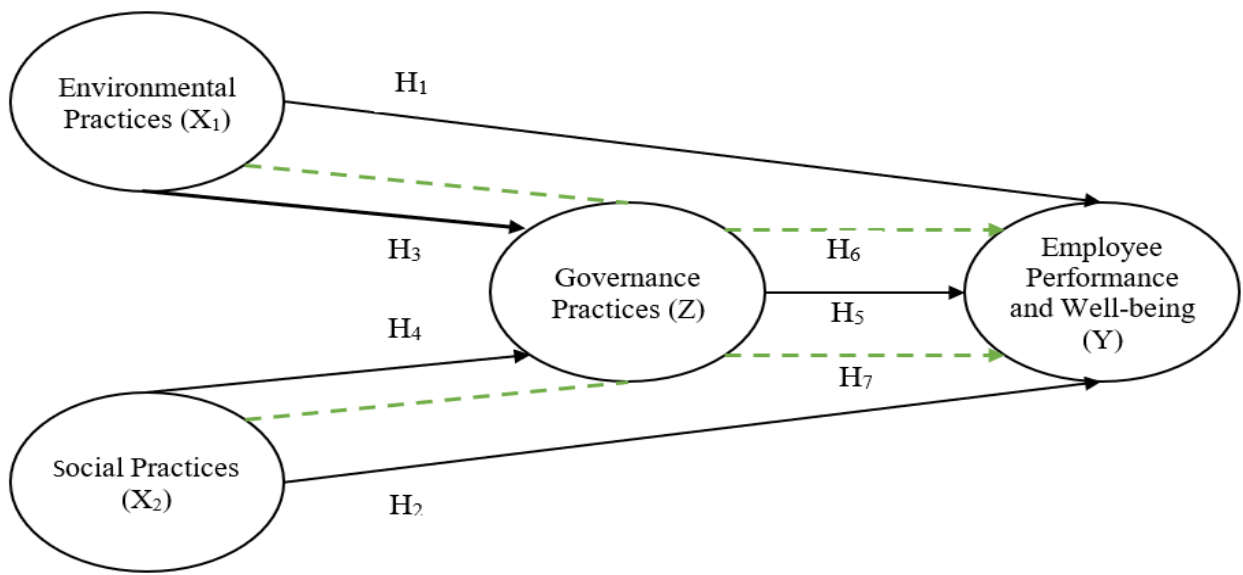

**Figure 1.** Conceptual Model

Through a systematic approach, this study framework helps analyze factors affecting employee efficiency and satisfaction. It provides a basis for formulating more sustainable and responsive policies to challenges in the higher education sector. The conceptual model of this research is illustrated in Figure 1. This framework is grounded in theoretical literature emphasizing the importance of synergy between sustainability practices and governance in supporting employee well-being and performance. Based on this conceptual framework, the following seven hypotheses are proposed:

**H$_1$:** Environmental practices have a direct positive effect on employee performance and well-being
**H$_2$:** Social practices have a direct positive effect on employee performance and well-being
**H$_3$:** Environmental practices positively influence governance practices
**H$_4$:** Social practices positively influence governance practices
**H$_5$:** Governance practices positively influence employee performance and well-being
**H$_6$:** Governance practices mediate the effect of environmental practices on employee performance and well-being
**H$_7$:** Governance practices mediate the effect of social practices on employee performance and well-being.

## 2. Research Methods

This study employs a quantitative method with a Partial Least Squares Structural Equation Modeling (PLS-SEM) approach to explore the relationships between environmental practices, social practices, governance practices, and employee performance and well-being at private universities in Bandung, West Java, Indonesia. The conceptual model applied in this research illustrates the relationships among the variables, with governance practices acting as a mediating variable. This approach is primarily focused on hypothesis development, as the study aims to test the relationships between these constructs based on established theories and empirical evidence in the literature. The development of hypotheses is driven by prior research on ESG practices and their impact on organizational outcomes. These are the foundations for testing the proposed relationships within the private university context.

PLS-SEM is particularly suited for this study because it handles complex models with multiple variables and intricate relationships between latent constructs. PLS-SEM is known for its flexibility in dealing with smaller sample sizes, non-normal data, and the ability to model both the measurement and structural components of the model simultaneously, which makes it a powerful tool for examining the specified relationships (Hair et al., 2021; Ringle et al., 2020). This technique allows for an in-depth analysis of the direct and indirect effects between the variables, providing insights crucial for understanding how ESG practices influence employee performance and well-being.

The research sample consists of 138 private universities in Bandung, with a total population of 828 individuals, including heads of various administrative departments. Stratified random sampling was used to ensure the proportional representativeness of each subgroup, resulting in 270 respondents. This sample size was determined using the Slovin formula with a 5% error margin, which is commonly used to achieve a reliable and representative sample for statistical analysis. The stratified sampling method ensures that various university administrative positions are adequately represented, contributing to a comprehensive understanding of how different roles perceive the impact of ESG practices on performance outcomes.

Data was collected through structured questionnaires distributed to the selected respondents. The questionnaires were designed to capture ESG practices, governance, and employee performance and well-being variables. The data collection process was rigorously structured to ensure accuracy and consistency, with measures to minimize response bias. After the data were collected, they were analyzed using SmartPLS 3.0, a widely used software for PLS-SEM analysis. SmartPLS 3.0 is chosen for its robustness in evaluating complex models with multiple latent variables and its ability to simultaneously assess measurement and structural models (Ghozali & Hengky, 2020).

During the data processing phase, construct validity was assessed using the Average Variance Extracted (AVE), while reliability was evaluated using Composite Reliability (CR) and Cronbach's Alpha. These validity and reliability tests are essential to ensure that the study's constructs are consistent and accurate representations of the theoretical variables. After meeting the validity and reliability criteria, structural model analysis was performed to evaluate the relationships between latent variables using path coefficients and R-squared values (Cheah et al., 2020). This approach provides a robust framework for testing the hypothesized relationships and offers insights into how ESG practices influence employee outcomes in the context of private universities in Bandung.

## 3. Results and Discussions

### 3.1. Measurement Model (Outer Model)

The measurement model, also known as the outer model, is used to evaluate the relationship between latent variables and their observed indicators. One of the key criteria in this evaluation is factor loading, which indicates how well each indicator represents the associated construct. A factor loading value of 0.70 or higher is generally considered acceptable, demonstrating a strong correlation between the indicator and the latent variable (Hair et al., 2021). Indicators with low loading values may be excluded to improve the model's validity and reliability. This step is crucial to ensure that each construct is measured accurately and meaningfully through its indicators. The results of the Measurement Model (Outer Model) can be seen in Figure 2 below.

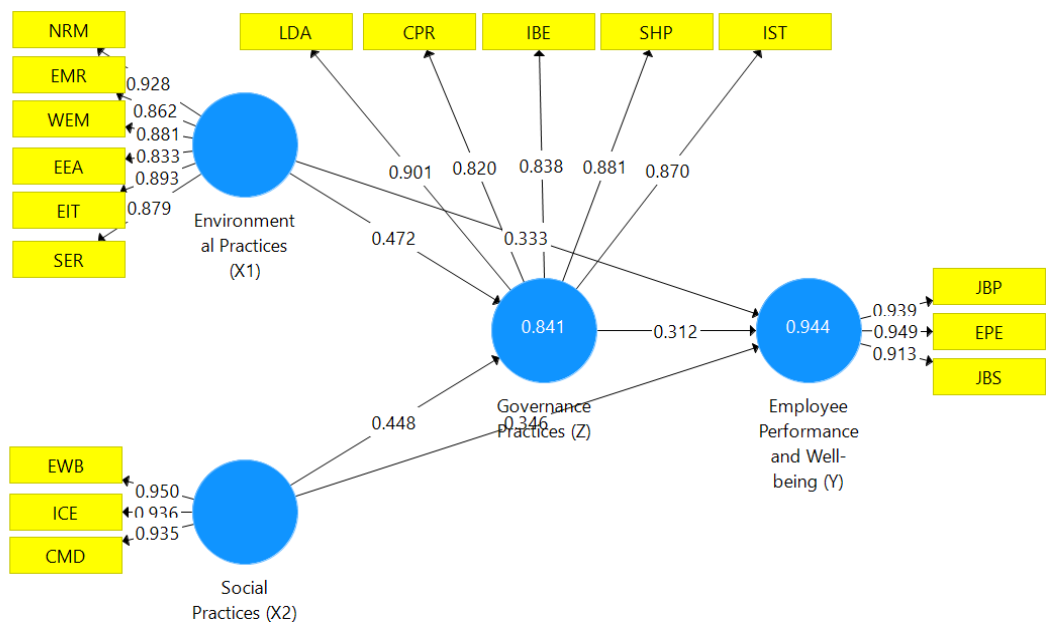

**Figure 2.** Outer Model

Figure 2 illustrates the outer model measurement for integrating ESG criteria in employee management. Each variable is constructed based on indicators representing sustainable human resource management practices. This model aims to explain the relationship between implementing ESG principles and improving employee productivity and quality of work. By applying this approach, it is expected to gain a more comprehensive understanding of the impact of ESG on employee behavior and performance within an organization.

Environmental practices are assessed using six indicators: Natural Resource Management (NRM) (0.928), Energy Management and Reduction and Reduction (EMR) (0.862), Waste Management and Reduction (WEM) (0.881), Environmental Education and Awareness (EEA) (0.833), Environmental Impact and Technology (EIT) (0.879), and Sustainable Environmental Reporting (SER) (0.893). These indicators demonstrate substantial contributions to measuring Environmental Practices. Social practices are evaluated through three indicators: Employee Well-being (EWB) (0.950), Inclusive and Collaborative Environment (ICE) (0.936), and Community Development (CMD) (0.935), which exhibit high factor loadings, indicating effective measurement in the ESG framework.

As a mediating variable, Governance practices are measured by six indicators: Leadership and Accountability (LDA) (0.901), Compliance and Regulation (CPR) (0.820), Integrity and Business Ethics (IBE) (0.838), Stakeholder Participation (SHP) (0.881), and Information System and Transparency (IST) (0.870), all showing significant contributions. The dependent variable, Employee Performance and Well-being (Y) is assessed through Job Performance (JBP) (0.939), Employee Engagement (EPE) (0.949), and Job Satisfaction (JBS) (0.913). All indicators have factor values above 0.7, demonstrating effective measurement in the ESG context (Cheah et al., 2020).

### 3.2. Test of Validity and Test of Reliability

Validity and reliability tests are essential steps in assessing the quality of research instruments. These tests are conducted through the Construct Reliability and Validity approach, including Composite Reliability (CR) and Average Variance Extracted (AVE) values. CR is used to ensure internal consistency among indicators within a construct, with an ideal value above 0.70, while AVE reflects the proportion of variance explained by the construct in relation to its indicators, with a recommended minimum value of 0.50 (Hair et al., 2021). A construct is considered valid and reliable when both values meet the criteria, indicating that the instrument consistently and accurately measures the intended construct. The results of the validity and reliability tests are presented in Table 1 below.

**Table 1.** Construct Reliability and Validity

|  | Cronbach's Alpha | rho_A | Composite Reliability | AVE |
|---|---|---|---|---|
| Employee Performance and Well-being (Y) | 0.927 | 0.927 | 0.953 | 0.872 |
| Environmental Practices ($X_1$) | 0.941 | 0.943 | 0.954 | 0.774 |
| Social Practices ($X_2$) | 0.935 | 0.935 | 0.958 | 0.884 |
| Governance Practices (Z) | 0.914 | 0.916 | 0.935 | 0.744 |

Source: processed data

Table 1 presents the results of the validity and reliability tests for the latent variables in this study. The findings indicate that all variables exhibit excellent validity and reliability. Specifically, Cronbach's Alpha values for all variables exceed 0.9, indicating high internal consistency (Hair et al., 2021). The Composite Reliability values for all variables are also greater than 0.9, confirming that these constructs are highly reliable. AVE values for each variable are greater than 0.7, demonstrating strong convergent validity, where the latent variables explain more than 50% of the variance from the construct indicators (Schuberth et al., 2023).

Specifically, the variable Employee Performance and Well-being shows an AVE value of 0.872, while environmental practices, social practices, and governance practices have AVE values of 0.774, 0.744, and 0.884, respectively. These values indicate that the indicators within these constructs are highly correlated with the measured latent variables. Therefore, it can be concluded that all constructs in this study possess excellent validity and reliability, allowing for further analysis with high confidence in the results obtained (Fornell & Larcker, 2016); (Marcoulides & Falk, 2018); (Hwang et al., 2021).

### 3.3. Structural Model Analysis (Inner Model)

The structural model, or inner model, examines the relationships between latent constructs and assesses the model's predictive power. Path coefficients and R-square ($R^2$) values are two key indicators in this analysis. Path coefficients reflect the strength and direction of the relationships between variables, where higher values indicate stronger relationships. Meanwhile, the R-square value indicates the proportion of variance in the dependent variable that can be explained by the independent variables, with values of 0.67, 0.33, and 0.19 considered substantial, moderate, and weak, respectively (Hair et al., 2021). These indicators are essential for validating the proposed hypotheses and evaluating the explanatory capability of the model. The results of Inner Model can be seen in Figure 3 below.

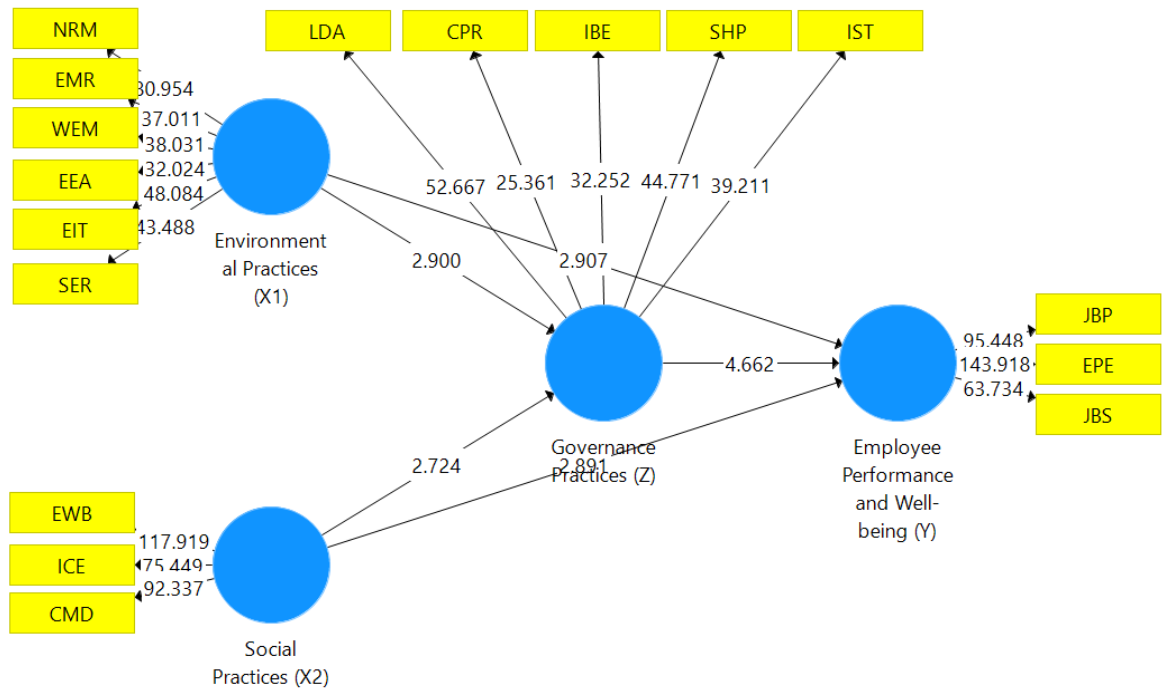

**Figure 3.** Inner Model

Figure 3 displays the inner model of the research on integrating ESG criteria in employee management. This model illustrates the causal relationships among latent variables. environmental practices impact governance practices and employee performance and well-being with path coefficients of 2.900 and 4.662, respectively, highlighting their significant role in enhancing governance and employee well-being. Social practices also influence governance practices and employee performance and well-being with path coefficients of 2.724 and 2.891, indicating their direct effect on improving employee outcomes. Governance practices mediate the relationship, showing a path coefficient of 4.662 with employee performance and well-being. Indicators for Governance practices have notable t-statistics: leadership and accountability (52.667), compliance and regulation (25.361), integrity and business ethics (32.252), stakeholder participation (44.771), and information system and transparency (39.211), demonstrating their significant contribution (cheah et al., 2020).

### 3.4. Hypothesis Testing

The general criteria for significance are that the t-value must be greater than 1.96, and the P-value must be less than 0.05. This method allows for the evaluation of the null hypothesis, which states that there is no effect, and the alternative hypothesis, which claims a significant relationship (Sugiyono, 2021). The results of this test will provide empirical evidence to support or reject the proposed hypotheses, thereby strengthening the validity of findings in the fields of sustainability and human resource management. The results of hypothesis testing can be seen in Table 2, which shows the path coefficients below.

**Table 2.** Path Coefficients

| | Original Sample (O) | Sample Mean (M) | Standard Deviation (STDEV) | t-statistics (\|O/STDEV\|) | P Values |
|---|---|---|---|---|---|
| Environmental Practices ($X_1$) => Employee Performance and Well-being (Y) | 0.333 | 0.322 | 0.114 | 2.907 | 0.004 |
| Social Practices ($X_2$) => Employee Performance and Well-being (Y) | 0.346 | 0.351 | 0.120 | 2.891 | 0.004 |
| Environmental Practices ($X_1$) => Governance Practices (Z) | 0.472 | 0.475 | 0.163 | 2.900 | 0.004 |
| Social Practices ($X_2$) => Governance Practices (Z) | 0.448 | 0.446 | 0.164 | 2.724 | 0.007 |
| Governance Practices (Z) => Employee Performance and Well-being (Y) | 0.312 | 0.317 | 0.067 | 4.662 | 0.000 |

Source: processed data

Table 2 above displays the path coefficients along with related statistics such as T-values and P-values to test the significance of these relationships. These statistical indicators help determine whether the data supports the hypothesized paths between variables. Generally, a path is considered statistically significant if the T-value exceeds 1.96 and the P-value below 0.05 (Hair et al., 2021). This analysis provides insights into the strength and direction of the influence each independent variable has on the dependent variable. The significance levels also indicate which ESG dimensions have the most substantial effect on employee outcomes.

### 3.5. The Influence of Environmental Practices on Employee Performance and Well-being

Environmental practices have a path coefficient of 0.333 with a T-value of 2.907 > 1.96 and a P-value of 0.004 < 0.05. This indicates that environmental practices positively and significantly impact employee performance and well-being. These findings align with the notion that organizations adopting sustainable environmental practices benefit in terms of social responsibility and experience improved employee productivity (Ahmed et al., 2020). Previous research suggests that a commitment to sustainability tends to enhance employee motivation, which, in turn, contributes to improved individual work performance (Tortia et al., 2022; Gyensare et al., 2024).

In the context of private universities in Bandung, the implementation of good environmental practices, such as efficient waste management and sustainable resource use, can create a more pleasant work environment and increase employee satisfaction. A study by (Marikyan et al., 2024) found that when employees in educational institutions perceive their organization as committed to sustainability, they are likely to have higher job satisfaction. This positive experience helps foster employee loyalty and improve staff retention (Alzadjali & Ahmad, 2024). Improvements in employee performance and well-being at private universities in Bandung can benefit the institution as a whole. Universities known for strong environmental practices are more attractive to prospective students and stakeholders, thus enhancing their market position (Yang & Xu, 2024). As students increasingly seek institutions that prioritize environmental issues, investing in these practices is not only regulatory compliance but also a strategic move for boosting organizational performance and ensuring long-term success. This aligns with the Resource-Based View (RBV) theory, which posits that an organization's resources, including its environmental practices, serve as valuable, rare, and inimitable assets that can enhance its competitive advantage (Gerhart & Feng, 2021). By focusing on sustainability, universities can leverage these resources to improve their institutional reputation and employee engagement, satisfaction, and productivity, all of which contribute to the institution's long-term success. The RBV underscores the idea that sustainable practices are strategic assets that align with stakeholder expectations, fostering both internal and external support and ultimately benefiting organizational outcomes.

### 3.6. The Influence of Social Practices on Employee Performance and Well-being

Social Practices ($X_2$) have a path coefficient of 0.346 with a T-value of 2.891 > 1.96 and a P-value of 0.004 < 0.05. These results indicate that good social practices also positively and significantly impact employee performance and well-being. This aligns with research suggesting that positive organizational social interactions can enhance employee well-being and productivity (Chtioui et al., 2023).

Social practices within an organization encompass various aspects of employee interaction, including team support, collaboration, and effective communication. Research by Dugan et al. (2023) shows that a work environment that supports positive social relationships boosts employee motivation and creates a conducive work atmosphere. At private universities in Bandung, the implementation of good social practices can foster more harmonious interactions between faculty, administrative staff, and students, as well as strengthen the sense of community among institution members. Additionally, increased solidarity and cooperation among team members can reduce conflicts and stress, which are crucial for maintaining employee well-being (Evans et al., 2024).

Implementing good social practices helps build a strong organizational culture where individuals feel valued and cared for. Research by Hill et al. (2024) shows that an inclusive and friendly work environment boosts job satisfaction and reduces staff turnover. At private universities in Bandung, when faculty and staff feel part of a supportive community, they are more committed and put in greater effort. Thus, effective social practices enhance both individual employee experiences and overall organizational performance. Developing healthy social relationships in educational institutions fosters a productive work environment, driving innovation and long-term success (Rajashekar & Jain, 2024). These findings are supported by social exchange theory, which posits that positive social interactions within an organization create a reciprocal relationship where employees are more likely to exhibit higher levels of commitment, job satisfaction, and performance (Blau, 2017). By fostering an inclusive and supportive environment, universities can enhance employee motivation and engagement, ultimately contributing to organizational growth and success.

### 3.7. The Influence of Environmental Practices on Governance Practices

The path coefficient between environmental practices and governance practices is 0.472 with a T-value of 2.900 > 1.96 and a P-value of 0.004 < 0.05. This indicates that environmental practices positively and significantly impact governance within the organization. This finding supports the view that sustainability efforts in environmental aspects are closely linked to improved transparency, accountability, and ethical standards in corporate governance. Sound environmental practices can strengthen effective governance in organizations (Bresciani et al., 2023). In the context of private universities in Bandung, the adoption of sustainable environmental practices is becoming increasingly important due to rising demands from various stakeholders. Universities that actively adopt environmental practices demonstrate a commitment to sustainability and enhance transparency and accountability in institutional governance. Research by Shah & Soomro (2023) shows that educational institutions implementing environmentally friendly policies tend to create better governance systems, thus meeting the expectations of students and community expectations. Furthermore, good environmental practices also provide a foundation for developing responsive and adaptive governance in educational organizations. When universities address environmental issues, they are better prepared to meet sustainability targets set by government and accreditation bodies (Habib & Mourad, 2024). This is consistent with findings by Hassan et al. (2024), which emphasize the importance of integrating environmental practices into governance frameworks to enhance the institution's presence and reputation in the educational market.

Thus, the positive relationship between environmental practices and governance practices indicates that focusing on sustainability not only improves organizational performance in terms of regulatory compliance but also fosters a culture oriented towards social responsibility in private universities in Bandung, ultimately contributing to strengthening the institution's position within the academic community and the broader society. These findings can be explained through the Institutional Theory, which suggests that organizations adopt practices perceived as legitimate and aligned with societal expectations. In this context, sustainability practices, including environmental and governance initiatives, help universities gain legitimacy, enhance their reputation, and maintain competitiveness in a socially conscious environment (Cockcroft, 2019).

### 3.8. The Influence of Social Practices on Governance Practices

Social practices have a path coefficient of 0.448 with a T-value of 2.724 > 1.96 and a P-value of 0.007 < 0.05. This indicates that social practices also positively and significantly impact governance. This result highlights the importance of fostering inclusive, fair, and supportive workplace environments in enhancing organizational governance. Good social practices within an organization can enhance the effectiveness of governance and decision-making (Sarhan & Al-Najjar, 2023). These practices may involve promoting employee well-being, ensuring diversity and equity, engaging in community development, and respecting human rights, which build trust, collaboration, and transparent communication across all organizational levels.

Implementing effective social practices is becoming increasingly crucial for creating a productive academic environment in private universities in Bandung. Research shows that institutions fostering positive social interactions among staff and students tend to have higher levels of engagement, contributing to more inclusive and participatory decision-making (Adeneye et al., 2023). By creating open discussion and collaboration space, universities can enhance team dynamics and add value to strategic decisions. Furthermore, good social practices help strengthen an inclusive organizational culture where every individual feels heard and valued. Research by Omaghomi et al. (2023) found that employee satisfaction and engagement increase when individuals have the opportunity to contribute to decision-making processes. This is particularly important in private universities, where faculty and administrative staff need to collaborate effectively to achieve the institution's vision and mission.

Moreover, the effectiveness of governance resulting from good social practices in universities can directly impact the institution's reputation with the public and prospective students. Research by Thuy et al. (2024) shows that organizations with a positive reputation related to sustainable social practices are better able to attract students and other stakeholders who value social responsibility and ethics. Therefore, implementing good social practices can strengthen the position of private universities in Bandung amidst the increasingly competitive educational market. This can be explained through the Social capital theory of Morrow & Scorgie-Porter (2017) and Chan (2018), which emphasizes the value of networks and relationships in creating a strong organizational reputation and fostering community engagement. By investing in social practices, universities build trust and cooperation, enhancing their competitive advantage in the education sector.

### 3.9. The Influence of Governance Practices on Employee Performance and Well-being

The path coefficient between Governance Practices and Employee Performance and Well-being is 0.312 with a T-value of 4.662 > 1.96 and a P-value of 0.000 < 0.05. This result indicates that governance practices have a positive and highly significant impact on employee performance and well-being. Good governance plays a crucial role in ensuring employee well-being and enhancing overall organizational performance (García-Juan et al., 2023).

Implementing effective governance practices is critical for managing resources and achieving academic and operational goals in private universities in Bandung. Research by Siverbo (2023) shows that institutions with good governance can enhance employee engagement and satisfaction, which directly impacts educational performance. This is particularly relevant as private universities face pressure to meet academic and social expectations from stakeholders, including students, faculty, and the community. Furthermore, transparent and accountable governance practices can build trust among employees and other stakeholders within the educational organization. Research by Jaiswal et al. (2024) reveals that institutions demonstrating a commitment to good governance have a higher reputation and can attract more students, especially in an era where prospective students are increasingly concerned about social responsibility and institutional ethics. Employee involvement in transparent decision-making processes can also strengthen their sense of belonging and commitment to the institution, thereby reducing staff turnover and improving overall performance (Jaiswal & Prabhakaran, 2024).

Implementing effective governance practices at private universities in Bandung contributes to employee performance and well-being and enhances the institution's ability to navigate challenges in the competitive educational market. Drawing from the Stakeholder Theory Freeman et al. (2021), which emphasizes the importance of creating value for various stakeholders (including employees, students, and the broader community), governance practices aligned with social and environmental responsibility can strengthen an institution's position. By addressing stakeholders' interests, universities are better equipped to respond to social and environmental changes, ensuring long-term sustainability and bolstering their reputation. This approach ultimately fosters a resilient institution that attracts and retains students, enhances staff engagement, and elevates its standing within the academic community.

**Table 3.** R Square

|  | R Square | R Square Adjusted |
|---|---|---|
| Employee Performance and Well-being (Y) | 0.944 | 0.943 |
| Governance Practices (Z) | 0.841 | 0.840 |

Source: processed data

According to Table 3, the R Square value for the variable employee performance and well-being (Y) is 0.944, and the Adjusted R Square value is 0.943, indicating that the independent variables in the model can explain 94.4% of the variability in employee performance and well-being. The R Square value governance practices is 0.841, and the Adjusted R Square value is 0.840, showing that environmental practices and social practices can explain 84.1% of the variability in governance practices. These results indicate that the model has strong predictive ability and significant relevance in the context of this research, supporting previous literature that highlights the importance of integrating environmental, social, and governance practices in improving organizational performance and employee well-being (Jain et al., 2024; Nugroho et al., 2024; Jorgji et al., 2024; Zhang et al., 2024; Habib & Mourad, 2024).

### 3.10. Testing Mediation Effects

Testing mediation effects is essential in understanding how one variable transmits its influence to another through an intermediary variable. This is typically assessed using total effects and specific indirect effects. The total effect represents the overall impact of an independent variable on a dependent variable, including both direct and indirect pathways, while specific indirect effects focus on the portion of the effect that is mediated through a specific variable (Hair et al., 2021). By evaluating these effects, researchers can identify whether and how mediation occurs within the model, offering insights into the underlying mechanisms of the relationships between variables. The results of testing mediation effects can be seen in Table 4 and 5.

**Table 4.** Total Effects

|  | Original Sample (O) | Sample Mean (M) | Standard Deviation (STDEV) | t-statistics (\|O/STDEV\|) | P Values |
|---|---|---|---|---|---|
| Environmental Practices ($X_1$) => Employee Performance and Well-being (Y) | 0.480 | 0.475 | 0.135 | 3.552 | 0.000 |
| Social Practices ($X_2$) => Employee Performance and Well-being (Y) | 0.486 | 0.491 | 0.136 | 3.564 | 0.000 |
| Environmental Practices (X1) => Governance Practices (Z) | 0.472 | 0.475 | 0.163 | 2.900 | 0.004 |
| Social Practices ($X_2$) => Governance Practices (Z) | 0.448 | 0.446 | 0.164 | 2.724 | 0.007 |
| Governance Practices (Z) => Employee Performance and Well-being (Y) | 0.312 | 0.317 | 0.067 | 4.662 | 0.000 |

Source: processed data

Table 4 above shows the total effects, which illustrate the total impact of the independent variables on the dependent variable, both directly and indirectly through the mediator (governance practices). Research by Hair et al. (2021) explains that a mediator can either strengthen or weaken the relationship between two variables, indicating that governance practices enhance the positive effects of environmental and social practices on employee performance and well-being. By examining mediation effects, researchers can identify the critical role of mediators in explaining the relationships between variables in the research model (Hwang et al., 2021).

In Table 5, it is shown that environmental practices have an indirect effect on employee performance and well-being through governance practices, with a coefficient value of 0.147, a T-value of 2.234, and a P-value of 0.026. This indicates environmental practices' positive and significant impact on employee performance and well-being through good governance. This suggests that when organizations implement effective environmental practices, they not only improve their sustainability outcomes but also foster better governance structures, which in turn enhance employee well-being and performance. Moreover, the results highlight that governance plays a crucial role as a mediator in translating environmental efforts into tangible benefits for employees, underlining the importance of integrating ESG criteria into organizational strategies. Thus, this finding further reinforces the idea that good governance is essential for maximizing the positive effects of environmental practices on overall employee outcomes.

**Table 5.** Specific Indirect Effects

|  | Original Sample (O) | Sample Mean (M) | Standard Deviation (STDEV) | t-sStatistics (\|O/STDEV\|) | P Values |
|---|---|---|---|---|---|
| Environmental Practices ($X_1$) => Governance Practices (Z) => Employee Performance and Well-being (Y) | 0.147 | 0.152 | 0.066 | 2.234 | 0.026 |
| Social Practices ($X_2$) => Governance Practices (Z) => Employee Performance and Well-being (Y) | 0.140 | 0.140 | 0.057 | 2.464 | 0.014 |

Source: processed data

Similarly, social practices show an indirect effect of 0.140, with a T-value of 2.464 and a P-value of 0.014. This suggests that social practices also significantly contribute to improving employee performance and well-being through the influence exerted by governance practices. In other words, the mediation effect test demonstrates that governance practices act as a link that strengthens the positive impact of environmental and social practices on employee performance outcomes.

### 3.11. The Influence of Environmental Practices on Employee Performance and Well-being through Governance Practices

The results from Table 5 indicate that the coefficient value for environmental practices affecting governance practices and subsequently contributing to employee performance and well-being is 0.147, with a T-value of 2.234 > 1.96 and a P-value of 0.026 < 0.05. This suggests that environmental practices significantly impact governance practices, which, in turn, contribute to the enhancement of employee performance and well-being. This means that good environmental practices can improve employee performance and well-being with the support of effective governance practices (Gyensare et al., 2024).

These findings align with research by Jain et al. (2024), which reveals that implementing good environmental practices enhances the organization's reputation and contributes to creating a better working environment, which, in turn, improves employee performance. Amidst the intense competition in the education market, private higher education institutions in Bandung must emphasize the importance of environmental practices in their governance to attract students and stakeholders concerned about sustainability.

Furthermore, Jaiswal et al. (2024) argue that solid governance practices enable the integration of environmental and social practices, creating a more supportive work environment. This aligns with the stakeholder theory (Freeman et al., 2021), which suggests that businesses and institutions should align their practices with the interests of all stakeholders to ensure mutual benefits. By effectively implementing environmental practices, supported by robust governance practices, private higher education institutions in Bandung can foster a work environment that enhances both individual employee performance and overall well-being. As these institutions face increasing social and environmental expectations, governance that emphasizes stakeholder engagement becomes crucial for adapting to these changes and maintaining long-term institutional success. This approach improves internal organizational dynamics and strengthens the institution's position in the broader educational market, meeting the evolving demands of students, faculty, and society.

### 3.12. The Influence of Social Practices on Employee Performance and Well-being through Governance Practices

Table 5 shows that the coefficient value for the path analysis between social practices and governance practices, which contributes to employee performance and well-being, is 0.140, with a T-statistic of 2.464 > 1.96 and a P-value of 0.014 < 0.05. This indicates that social practices significantly influence the improvement of employee performance and well-being through good governance. Previous research shows that the implementation of effective social practices can enhance job satisfaction and overall individual productivity (Gyensare et al., 2024; Aubouin-Bonnaventure et al., 2024).

Increasing social interaction and collaboration between teaching staff and administration can create a more harmonious work environment in private higher education institutions in Bandung. Research by Evans et al. (2024) shows that positive social relationships among employees in higher education institutions directly contribute to increased engagement and job satisfaction. Furthermore, effective social practices within these institutions can also support the implementation of responsive and transparent governance policies, which are increasingly valued by students and the community (Ajayi & Udeh, 2024). Social exchange theory can explain this, which posits that positive social interactions, characterized by trust and reciprocity, lead to mutual benefits for both individuals and organizations (Blau, 2017)When social practices foster a sense of belonging and fairness, they enhance governance transparency and responsiveness, improving both employee morale and institutional reputation. Therefore, enhancing social practices, supported by good governance, will create a productive work environment, promote employee engagement, and support the development of employee performance in private higher education institutions in Bandung. By prioritizing these practices, universities can better meet stakeholder expectations and navigate the challenges of an increasingly competitive educational market.

### 4.  Conclusions

The study results show that ESG significantly influence employee performance and well-being in private higher education institutions in Bandung. Effective environmental practices enhance employee productivity and well-being, while social practices foster positive relationships and job satisfaction. These findings are crucial for institutions aiming to stay competitive in the growing education market, as adopting ESG principles meets societal expectations and boosts institutional reputation. Institutions should establish clear frameworks for sustainable practices, such as mandatory ESG training for staff and creating a task force to oversee ESG implementation. This will integrate ESG principles into daily operations and long-term strategy, ensuring sustainability and organizational excellence. Therefore, private higher education institutions in Bandung should prioritize sustainability practices, involving employees in decision-making to improve performance and contribute to broader social and environmental goals.